\documentclass[aps,prl,preprintnumbers,amsmath,amssymb,latexsym,nofootinbib,array,enumerate,letter,twocolumn,superscriptaddress]{revtex4-1}

\usepackage{here}
\usepackage{comment,braket}
\usepackage{epsf}
\usepackage{amsmath,amssymb, mathrsfs, physics,}
\usepackage{graphics}
\usepackage{amsfonts}
\usepackage{amssymb}
\usepackage{latexsym}
\usepackage{color}
\usepackage{ulem}
\input{colordvi.tex}
\usepackage{feynmp}
\usepackage{balance}

\usepackage[pdftex]{graphicx}
\usepackage{bm}
\usepackage[pdftex]{hyperref}
\usepackage[svgnames]{xcolor}
\definecolor{phthaloblue}{rgb}{0.0, 0.06, 0.54}
\hypersetup{
    colorlinks=true,
    linkcolor=phthaloblue,
    citecolor=blue,
    filecolor=blue,
    urlcolor=phthaloblue,
    }

\newcommand{\CONCEPT}{\textsc{co\textsl{n}cept}}
\newcommand{\nc}{\newcommand}
\nc{\beq}{\begin{equation}}
\nc{\eeq}{\end{equation}}
\newcommand{\arh}{{a_{\mathrm{eq}}}}
\newcommand{\Hrh}{{H_{\mathrm{eq}}}}
\newcommand{\As}{A_{\mathrm{s}}}

\usepackage{tabularx}
\usepackage{array}
\newcolumntype{P}[1]{>{\centering\arraybackslash}p{#1}}
\newcommand{\ra}[1]{\renewcommand{\arraystretch}{#1}}

\makeatletter
\def\hlinewd#1{%
\noalign{\ifnum0=`}\fi\hrule \@height #1 \futurelet
\reserved@a\@xhline}

\renewcommand\onecolumngrid{% <<<<<<
\do@columngrid{one}{\@ne}%
\def\set@footnotewidth{\onecolumngrid}% <<<<<<<<<<<<<<<<
\def\footnoterule{\kern-6pt\hrule width 1.5in\kern6pt}%
}

\renewcommand\twocolumngrid{% <<<<<<
        \def\footnoterule{% restore rule
        \dimen@\skip\footins\divide\dimen@\thr@@
        \kern-\dimen@\hrule width.5in\kern\dimen@}
        \do@columngrid{mlt}{\tw@}
}%

\makeatother
\newcolumntype{L}[1]{>{\raggedright\let\newline\\\arraybackslash\hspace{0pt}}m{#1}}
\newcolumntype{C}[1]{>{\centering\let\newline\\\arraybackslash\hspace{0pt}}m{#1}}
\newcolumntype{R}[1]{>{\raggedleft\let\newline\\\arraybackslash\hspace{0pt}}m{#1}}
\renewcommand{\arraystretch}{1.2}

\begin{document}

%%\preprint{}

\title{Stochastic Gravitational Waves from Early Structure Formation}

\author{Nicolas Fernandez}
\affiliation{NHETC, Department of Physics and Astronomy, Rutgers University, Piscataway, NJ 08854, USA}

\author{Joshua W.~Foster}
\email{jwfoster@mit.edu}
\affiliation{Center for Theoretical Physics, Massachusetts Institute of Technology, Cambridge, MA 02139, USA}

\author{Benjamin Lillard}
\affiliation{Department of Physics, University of Oregon, Eugene, OR 97403, USA}

\author{Jessie Shelton}
\affiliation{Center for Theoretical Physics, Massachusetts Institute of Technology, Cambridge, MA 02139, USA}
\affiliation{Illinois Center for Advanced Studies of the Universe, University of Illinois at Urbana-Champaign, Urbana, IL 61801, USA}
\date{\today}
\preprint{MIT-CTP/5660}

\begin{abstract}
Early matter-dominated eras (EMDEs) are a natural feature arising in many models of the early universe and can generate a stochastic gravitational wave background (SGWB) during the transition from an EMDE to the radiation-dominated universe required by the time of Big Bang Nucleosynthesis. 
While there are calculations of the SGWB generated in the linear regime, no detailed study has been made of the nonlinear regime.
We perform the first comprehensive calculation of GW production in EMDEs that are long enough that density contrasts grow to exceed unity, using 2a hybrid $N$-body and lattice simulation to study GW production from both a metastable matter species and the radiation produced in its decay. 
We find that nonlinearities significantly enhance GW production up to frequencies at least as large as the inverse light-crossing time of the largest halos that form prior to reheating. The resulting SGWB is within future observational reach for curvature perturbations as small as those probed in the cosmic microwave background, depending on the reheating temperature. 
Out-of-equilibrium dynamics could further boost the induced SGWB, while a fully relativistic gravitational treatment is required to resolve the spectrum at even higher frequencies.
\end{abstract}

\maketitle

\noindent {\bf Introduction.}---%
Cosmologically generated gravitational wave (GW) backgrounds provide a unique opportunity to study the state of the very early universe at temperatures above big bang nucleosynthesis (BBN) and represent an important new physics target for a number of upcoming and proposed observatories, such as LISA \cite{LISA:2017pwj,LISACosmologyWorkingGroup:2022jok}, DECIGO \cite{Kawamura:2006up,Kawamura:2020pcg}, BBO \cite{Harry:2006fi}, $\mu$Ares \cite{Sesana:2019vho}, pulsar timing with SKA \cite{Janssen:2014dka}, and others. A notably minimal scenario in which a stochastic GW background (SGWB) may be generated is through structure formation during an early matter-dominated era (EMDE). EMDEs arise in a wide variety of well-motivated contexts \cite{Allahverdi:2020bys}, making early structure formation a potentially powerful probe of the particle physics of post-inflationary reheating \cite{Easther:2010mr,Jedamzik:2010dq,Jedamzik:2010hq, Erickcek:2011us, Barenboim:2013gya, Fan:2014zua,Amin:2014eta,Amin:2019ums, Martin:2019nuw, Musoke:2019ima, Niemeyer:2019gab, Eggemeier:2020zeg, Eggemeier:2021smj, Papanikolaou:2022chm,Lozanov:2022yoy, Eggemeier:2023nyu}, secluded dark sectors \cite{Zhang:2015era,Blanco:2019eij,Dror:2017gjq,Erickcek:2020wzd,Erickcek:2021fsu}, and natural axion dark matter models \cite{Visinelli:2009kt,Nelson:2018via,Blinov:2019jqc}.  Observationally, an EMDE prior to BBN is consistent with all cosmological observation so long as the return to radiation domination (RD) occurs at a reheating temperature $T_{\rm{eq}} \gtrsim \mathrm{few}\,\mathrm{MeV}$ \cite{deSalas:2015glj,Hasegawa:2019jsa}. 

The growth of scalar perturbations during an EMDE induces stochastic GWs at second order in cosmological perturbation theory; see \cite{Domenech:2021ztg} for a review.  
Provided the density contrasts $\delta$ remain within the perturbative regime, the induced GW signal can be studied using standard tools from cosmological perturbation theory.
This ``linear'' GW spectrum depends quadratically on the amplitude of the primordial curvature power spectrum, $\As$, and is maximized for modes that are horizon-size at reheating \cite{Assadullahi:2009nf, Kohri:2018awv, Inomata:2019zqy,  Pearce:2023kxp}. For this linear GW spectrum to be observable, either the metastable species responsible for realizing the EMDE must undergo faster than exponential decays \cite{Inomata:2020lmk, Pearce:2023kxp}, or $\As$ must be significantly enhanced on small scales even in the most optimistic sensitivity scenarios.  For marginally detectable $\As \gtrsim 10^{-6}$, this limits the duration of the EMDE to at most seven $e$-folds of scale factor growth before nonlinear structure forms.

\begin{figure*}[!ht]
\includegraphics[width=0.9\textwidth]{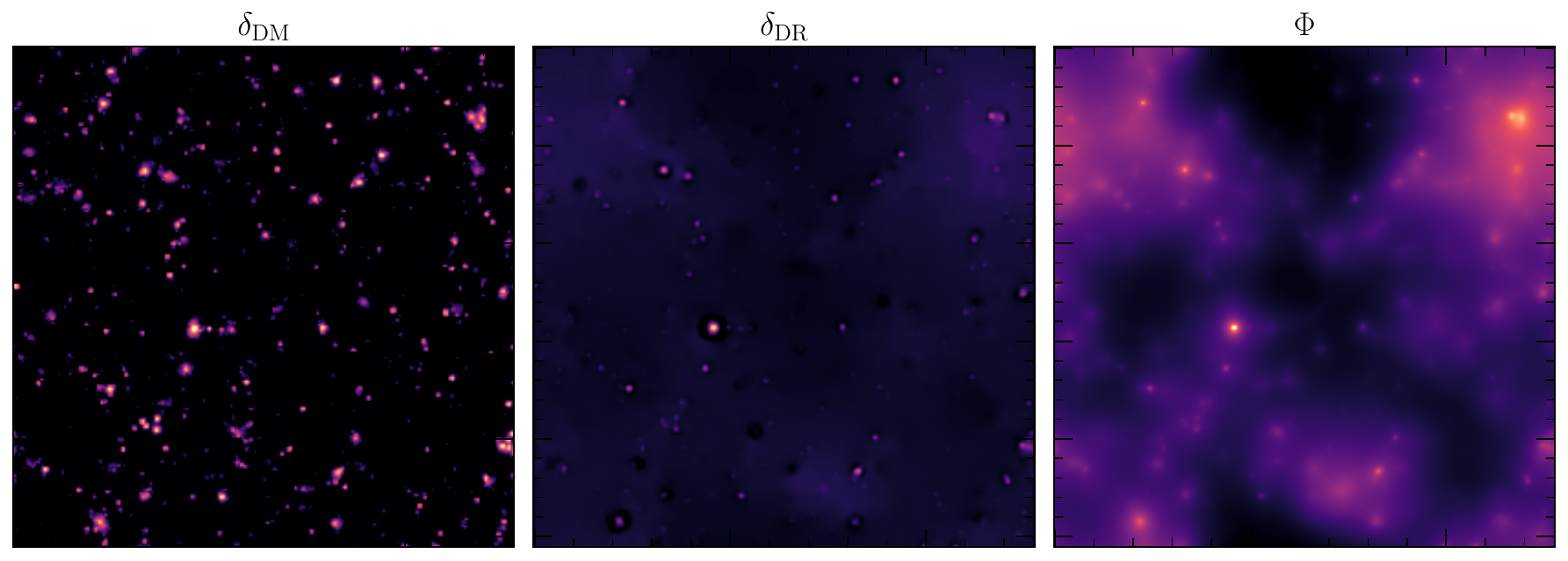}
\caption{(\textit{Left}) The density contrast field of DM obtained by meshing particle data at $a_\mathrm{eq}$ from the $\mathtt{S_{mid}}$ simulation. DM clusters, realizing overdensities that have grown as large as $\delta \approx 2300$. (\textit{Middle}) The density contrast field of DR, which does not cluster but traces the spatial distribution of the DM that sources it. In this simulation, we observe overdensities as large as $\delta = 10$. (\textit{Right}) The Newtonian scalar potential produced by the sum of the DM and DR energy densities.}
\label{fig:DensitySlice}
\vspace{-3ex}
\end{figure*}

Moreover, it has been suggested that nonlinear dynamics in the large-$\delta$ regime may amplify the GW production to more readily observable levels \cite{Jedamzik:2010hq}. An accurate calculation of the induced SGWBs produced from nonlinearities during an EMDE is critical in order to accurately characterize the signal associated with this scenario.
A recent effort \cite{Eggemeier:2022gyo} aimed to simulate and quantify the GW production from long EMDEs that realize a nonlinear matter density field prior to decay via an $N$-body simulation approach. In this \textit{Letter}, we present work that builds considerably on this first calculation.
We perform larger simulations that evolve the metastable decaying matter (DM) species as well as the radiation it produces via decay with fully consistent equations of state and background cosmological evolution.  Incorporating the decay radiation (DR) ensures that we capture not only the details of halo formation during the EMDE but also the subsequent halo evaporation and radiation emission during and after the reheating process. Coupled into our cosmological $N$-body simulations is a real-time evaluation of the induced tensor perturbations, fully accounting for nonrelativistic matter, relativistic radiation, and scalar potential sources. 

{\bf Induced Tensor Perturbations.}---%
Tensor perturbations to a Friedmann-Robertson-Walker metric in conformal Newtonian gauge can be numerically evolved by
\begin{equation}
    h_{ij}'' + 2 \mathcal{H} h_{ij}' - \nabla^2 h_{ij} = 4 \mathcal{S}_{ij}^\mathrm{TT} ,
    \label{Eq:EoM}
\end{equation}
where primes indicate differentiation with respect to conformal time, $\mathcal{H}$ is the conformal Hubble parameter, $\nabla^2$ is the Laplacian evaluated with respect to comoving coordinates, and $\mathcal{S}_{ij}^{TT}$ is the transverse-traceless (TT) component of a source tensor. The source tensor relevant for tensor mode production is given by
\begin{equation}
    \mathcal{S}_{ij} = 8 \pi G a^2 \mathcal{T}_{ij} -4 \Phi \partial_i \partial_j \Phi -2  \partial_i \Phi\partial_j \Phi,
    \label{eq:SourceTensor}
\end{equation}
where $\mathcal{T}_{ij}$ is the stress-energy tensor and $\Phi$ is the first-order Bardeen potential \cite{Ananda:2006af, Baumann:2007zm, Ali:2020sfw}. The $TT$ component can be extracted via spectral projection as developed in \cite{Garcia-Bellido:2007fiu, Dufaux:2007pt}. A key point here is that our source tensor is composed of quadratic combinations of the scalar potential $\Phi$ together with the full nonlinear stress-energy tensor for both radiation and matter species. 

Tensor modes are continuously sourced during matter domination, which complicates the identification of the energy density in propagating GWs during the EMDE itself.  After the EMDE ends, however, the sources decay and the tensor modes may be taken to be freely propagating, rendering the identification of the final energy density in GWs straightforward \cite{Hwang:2017oxa, Ali:2020sfw,Domenech:2020xin}. 

{\bf Simulation Framework.}---%
To evaluate tensor mode production following Eq.~\ref{Eq:EoM}, we must evaluate the gravitational dynamics of DM and DR through an EMDE and subsequent transtion to RD.  We do so using a modified version of the massively parallel code \CONCEPT\ \cite{Dakin:2021ivb}. \CONCEPT\ is capable of both \textit{N}-body  and relativistic fluid dynamics treatments of matter and radiation with non-trivial equation of state \cite{Dakin:2017idt, Tram:2018znz, Dakin:2019dxu, Dakin:2019vnj}. We note that \CONCEPT\ performs Newtonian simulations, meaning that the scalar potential propagates instantaneously, unlike our tensor perturbations, which propagate merely at $c$. As a result, our simulations realize unphysically rapid dynamics at the smallest scales that must be interpreted carefully. 

Using custom-generated initial conditions in the conformal Newtonian gauge associated with a long EMDE, we use \CONCEPT\ to evaluate the gravitational evolution of DM as a particle species as it simultaneously decays to produce a DR fluid.
We consider an initially scale-invariant spectrum of adiabatic perturbations, and study three different values of the amplitude $\As$ of the curvature power spectrum as described further below.
The decay of DM into DR proceeds at a constant rate $\Gamma \sim H_{\mathrm{eq}}$. Perturbative analyses of this system have been done in (\textit{e.g.}) \cite{Ichiki:2004vi, Erickcek:2011us,Audren:2014bca,Poulin:2016nat}. 

The radiation both fully contributes to and experiences the gravitational dynamics of our simulation; we model it by assumption as a perfect fluid. This assumption is line with the treatment of, \textit{e.g.}, \cite{Assadullahi:2009nf, Kohri:2018awv, Inomata:2019zqy}, and lets us compare directly to the perturbative calculation of Ref.~\cite{Inomata:2019zqy} in the linear regime. 

Our simulations begin at $a_i \approx 10^{-5}$ in the matter-dominated era, advancing through matter-radiation equality at $a_{\rm{eq}} \approx 0.2$ and ending deep in the radiation-dominated era at $a_f \equiv 1$. At $a_i$, density contrasts are sufficiently small that initial conditions can be generated with the Zel'dovich approximation \cite{1970A&A.....5...84Z}. Our simulations end late, with a return to RD realized through an adiabatically-evolving particle mass for the DM and a self-consistently evaluated background expansion rate, as in \cite{Dakin:2019dxu}, together with a real-space treatment inhomogeneously sourcing the DR following the DM distribution, which we have implemented here for the first time.  As a result, our simulations capture the effects of not just the collapse of gravitationally-bound DM structures as studied in \cite{Eggemeier:2022gyo}, but also the GW emission from the radiation produced in DM decay, which was conjectured to be the dominant source in the nonlinear regime \cite{Jedamzik:2010hq}. As an illustration of data evolved by our simulation, we present 2D sliceplots of DM densities, DR densities, and the scalar potential in Fig.~\ref{fig:DensitySlice}.

{\bf Gravitational Wave Calculation.}---%
We calculate GW production in the nonlinear EMD-to-RD scenario in real time as part of our simulations following Eq.~\ref{Eq:EoM}. Radiation contributions to the source tensor are calculated directly from the fluid grids evolved by \CONCEPT. Meshes of the DM energy and momentum densities, which are used to source the DR, and meshes of the DM stress-energy tensor elements, which contribute to the source for tensor perturbations, are generated from particle data using a deconvolved piecewise cubic spline interpolation \cite{Sefusatti:2015aex}.
Meanwhile, the Bardeen potential is calculated from the total matter and radiation density in the Newtonian approximation, which is sufficient as our initial conditions provide power only to modes that are subhorizon at the decay time when the GW signal is generated. 

We perform simulations in a comoving box with periodic boundaries. Fluid dynamics are evaluated with a Kurganov-Tadmor scheme using a second-order forward Runge-Kutta with a van~Leer flux limiter \cite{KURGANOV2000241}.
We evolve   Eq.~\ref{Eq:EoM} in nondissipative special form \cite{Figueroa:2020rrl} using a St\"{o}rmer-Cowell linear multistep method with a 5$^\mathrm{th}$ order forward-step and $6^\mathrm{th}$ order backward-step in Predict-Evaluate-Correct-Evaluate mode \cite{doi:https://doi.org/10.1002/9781119121534.ch4}. All spatial differentiations are performed spectrally, and our timestep satisfies the Courant–Friedrichs–Lewy condition with $\Delta \tau = \Delta x / 20$, where $\Delta x$ is the comoving lattice scale.

\begin{table}[!t]{
    \ra{1.2}
    \begin{center}
    \tabcolsep=0.07cm
    \begin{tabular}{c*{4}{C{0.085\textwidth}}}
    \hlinewd{1.5pt} 
    Sim. Name & $L_\mathrm{box} / \tau_{\mathrm{eq}}$& $N$ & $\As$ &  $k_\mathrm{nyq.}/k_\mathrm{NL}$ \\ \hlinewd{1.5pt}
    $\mathtt{Linear}$ & 3.84 & $560^3$ & $5.4 \times 10^{-13}$ & -- \\ \hlinewd{1.0pt}
    $\mathtt{S_{low}}$ & 3.84 & $560^3$ & $5.4 \times 10^{-5}$ & 11 \\ \hlinewd{1.0pt}
    $\mathtt{S_{mid}}$ & 0.88 & $256^3$ & $5.4 \times 10^{-5}$ & 22 \\ \hlinewd{1.0pt}
    $\mathtt{S_{hi}}$ & 0.96 & $560^3$ & $5.4 \times 10^{-5}$ & 44 \\ \hlinewd{1.0pt}
    $\mathtt{W_{low}}$ & 3.84 & $840^3$ & $1 \times 10^{-5}$ & 11 \\ \hlinewd{1.0pt}
    $\mathtt{W_{mid}}$ & 0.88 & $384^3$ & $1 \times 10^{-5}$ & 22 \\ \hlinewd{1.0pt}
    \end{tabular}\end{center}}
\caption{A summary of the six simulations presented in this work. We provide the box volume measured in terms of the conformal time at matter-radiation equality, the resolution parameter $N$, the choice of $\As$ for the scale-invariant initial curvature perturbation, and the small-scale resolution figure of merit $k_\mathrm{nyq.}/k_\mathrm{NL}$. See text for details. }
\label{tab:SimSpecs}
\vspace{-3ex}
\end{table}

\begin{figure*}[!ht]
\includegraphics[width=0.99\textwidth]{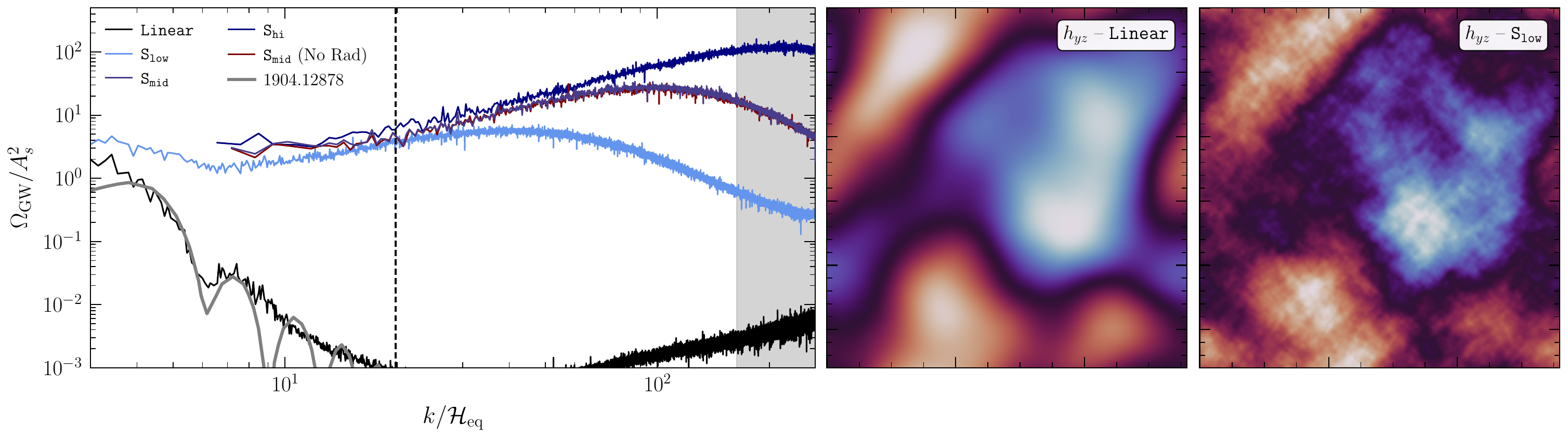}
\caption{(\textit{Left.}) The spectrum of GW energy density in the $\mathtt{Linear}$ simulation and the strongly nonlinear simulation at the three resolution choices considered in this work in light, medium, and dark blue in order of ascending small-scale resolution. In red, we show the result from an $\mathtt{S_{mid}}$ simulation performed without DR, which achieves a GW spectrum identical to that of the $\mathtt{S_{mid}}$ simulation that includes DR. The dashed black line indicates $k_\mathrm{NL}$ while the grey band indicates frequencies above the light-crossing time of the largest halos. (\textit{Right Panels.}) 2D slices of the $h_{yz}$ metric perturbation field evaluated from our $\mathtt{Linear}$ and $\mathtt{S_{low}}$ simulations, which use identical box volume, resolution, and timestepping, differing only in terms of the value of $\As$. Considerably enhanced small-scale structure can be observed in the $\mathtt{S_{low}}$ simulation, which realizes nonlinear gravitational dynamics within the matter species.}
\label{fig:GWSlice}
\vspace{-3ex}
\end{figure*}

Tab.~\ref{tab:SimSpecs} summarizes the simulations considered in this work. Our simulations are specified by a box dimension $L_\mathrm{box}$, a resolution parameter $N$ where $N^3$ is both the number of lattice sites for GW evolution and the number of particles, and $\As$. We summarize our small-scale resolution in terms of $k_\mathrm{nyq.}/k_\mathrm{NL}$ where $k_\mathrm{nyq.}$ is the Nyquist frequency of the mesh and $k_\mathrm{NL}$ is the smallest $k$ for which $\Delta_k^2 \geq 1$ at matter-radiation equality. We perform three strongly nonlinear simulations at $\As = 5.4 \times 10^{-5}$, labeled as $\mathtt{S_{low}}$, $\mathtt{S_{mid}}$, and $\mathtt{S_{hi}}$ in terms of their small-scale resolution, and two weakly nonlinear simulations at $\As = 10^{-5}$, labeled $\mathtt{W_{low}}$ and $\mathtt{W_{mid}}$. We also perform a simulation ($\mathtt{Linear}$) with $\As = 5.4\times 10^{-13}$, which does not realize any nonlinearity within the simulation resolution and enables comparison with \cite{Inomata:2019zqy}. In Fig.~\ref{fig:GWSlice}, we depict 2D slices of the metric perturbations $h_{ij}$ evaluated from two simulations at $a_\mathrm{eq}$. %The runtime of a simulation at fixed volume scales at least as $N^{4/3} \ln N$, making more resolved simulations prohibitively expensive.

{\bf Results.}---%
The total GW energy density is calculated by $\rho_\mathrm{GW} = \langle h_{ij}' h'^{ij} \rangle/(32 \pi G a^2)$,
with associated energy-density spectrum 
$\Omega_\mathrm{GW} = 1/\rho_\mathrm{c} \;  d \rho_\mathrm{GW}/d \ln k $.
We calculate this spectrum from the late-time field configuration of the tensor perturbations in our simulations. In the left panel of Fig.~\ref{fig:GWSlice}, we present the energy-density spectrum of GWs generated in our strongly nonlinear simulations. We also compare our linear simulation to the perturbative calculation developed in \cite{Inomata:2019zqy}, which used approximated fitting formulae that do not capture the full oscillatory behavior of the time-evolving scalar metric perturbations that enter the source for tensor modes. The results of our linear regime simulation demonstrate relatively good agreement with the perturbative calculation in peak height, location, and fall-off. In our nonlinear regime simulations, while we do not find significant enhancement of power at the reheat-scale peak, the power at smaller scales is considerably enhanced.

One of our main results is that our DR is not an important source of GW emission in the nonlinear regime.  This is shown in  Fig.~\ref{fig:GWSlice}, where simulations with and without radiation are shown to realize an identical GW spectrum.
On linear scales, our results demonstrate that structure formation does not result in substantial enhancement of the GW spectrum over the perturbative prediction. 
Since we find that the radiation fluid sourced by small-scale nonlinearities is inefficient at transporting anisotropic small-scale power to larger scales, the dominant source for GWs on horizon scales remains the decay of the scalar potentials. Examining Eq.~\ref{Eq:EoM} and Eq.~\ref{eq:SourceTensor} reveals that source terms quadratic in $\Phi$ have the effect of transferring power across scales down to the horizon scale at reheating, where GW production is efficient. However, in the nonlinear regime, the growth of overdensities slows, leading to a decrease of the scalar potential relative to the linear theory prediction. As a result, matter nonlinearities result in a scale-dependent suppression of scalar potentials and the GWs they produce relative to the linear-theory prediction, which was noted by \cite{Assadullahi:2009nf}. 

On the other hand, we observe significant GW emission at frequencies corresponding to scales that have gone nonlinear.  On these scales, the $TT$ source tensor receives its dominant contribution from DM. The collapse into gravitationally-bound structures generates anisotropic stress during shell-crossing, efficiently sourcing GWs at frequencies associated with the timescales on which DM halos evolve.

\begin{figure*}[!ht]
\includegraphics[width=0.99\textwidth]{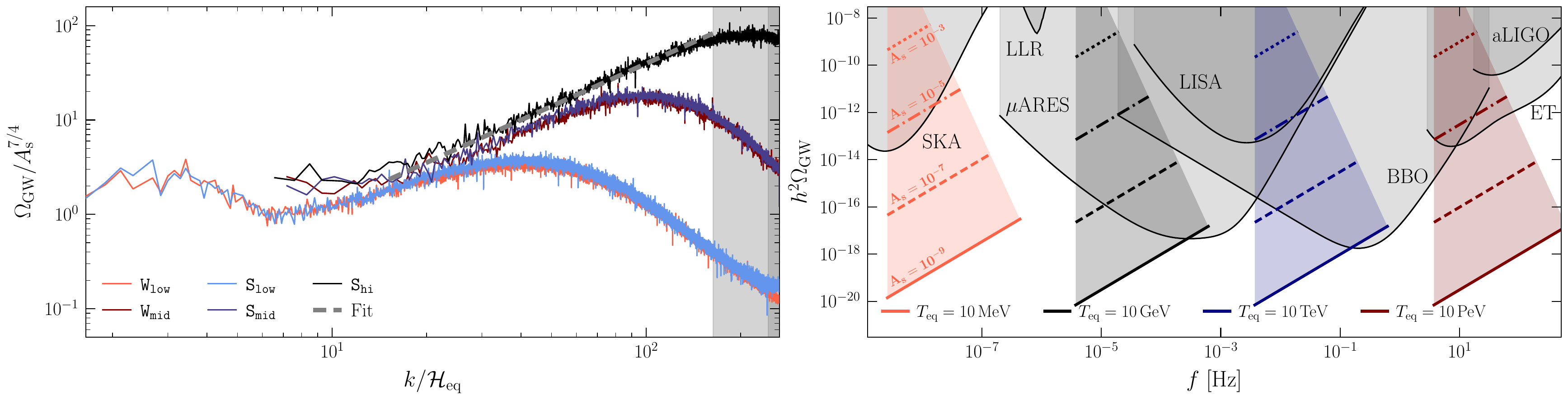}
\caption{(\textit{Left.}) A comparison of GW spectra between our strongly and weakly nonlinear simulations at equal low- and mid-resolution of the small-scale dynamics, %for the low-resolution and mid-resolution scenarios, 
demonstrating that the morphology of the GW spectrum is independent of $\As$ while its amplitude scales as $\As^{7/4}$. The light (dark) grey band indicates frequencies above the light-crossing time of the largest virialized halos in the strongly (weakly) nonlinear simulations. The dashed grey line shows the fit of Eq.~\ref{eq:Spec}. (\textit{Right.}) Predictions for the induced GW spectrum from Eq.~\ref{eq:Spec} for some representative choices of $T_\mathrm{eq}$ and $\As$, compared to projected observational sensitivities for future GW observatories obtained from \cite{Campeti:2020xwn, Blas:2021mqw, Blas:2021mpc}.}
\label{fig:AsDep_observability}
\vspace{-3ex}
\end{figure*}

The GW emission from collapsing halos is dominated by the largest halos for two reasons. First, the stress-energy associated with a halo is given parametrically by  $M_h v_h^2$ where $M_h$ is the mass of the halo and $v_h \propto M_h^{1/2}$ the typical speed of its constituents. Second, the largest halos are the latest-forming, and so their GW emission is least diluted by redshifting prior to the return to RD, as discussed in \cite{Jedamzik:2010hq,Eggemeier:2022gyo}.  Thus the amplitude of the emitted GW spectrum is set by the typical mass scale of halos that are collapsing immediately prior to reheating. Using the Press-Schechter mass function, we can estimate the typical latest-forming halo mass as well as the number density of such halos at reheating in terms of $\As$ and $\mathcal{H}_\mathrm{eq}$.  Assuming a collapsing halo contributes to the GW source tensor in an amount proportional to its total energy, this simple estimate then predicts that the GW energy density sourced by collapsing halos should scale as $\Omega_{\mathrm{GW}} \propto  \As^{7/4}$, independent of  $\mathcal{H}_\mathrm{eq}$ (see Supplemental Material). We observe exactly this scaling in Fig.~\ref{fig:AsDep_observability}.

However, the frequency spectrum depends on the dynamical timescale of these halos, which is determined by their density, not their mass. This density is in turn set by the background matter density at the time of formation. Thus the frequency spectrum is determined by the reheating timescale and is independent of the mass scale $M_h$. This is shown in the left panel of Fig.~\ref{fig:AsDep_observability}. The cutoff we observe in the small-scale GW power is not physical: as we see from varying $N$ across $\mathtt{S_{low}}$, $\mathtt{S_{mid}}$, and $\mathtt{S_{hi}}$, it comes from the resolution of the simulation. 

However, we caution that the Newtonian $N$-body model allows collapsing halos to act as coherent sources for GWs with frequencies substantially above the halo's light-crossing time, and thus for frequencies above the inverse light-crossing time of the largest halos in our simulations, the GW spectrum should be interpreted with care. We expect that, in a more causal treatment, the power radiated by a halo on scales smaller than its light-crossing time would be suppressed by the incoherence of emission across its volume, leading to an $M_h$ and therefore $\As$-dependent cutoff.

Comparing to our highest resolution simulation $\mathtt{S_{hi}}$ at $k$ below its resolution cutoff, we find our late-time GW power spectrum to be well-approximated by the power-law fit
\begin{equation}
\label{eq:Spec}
\Omega_\mathrm{GW}(k) \approx 0.05 \times \As^{7/4} \left( \frac{k}{\mathcal{H}_\mathrm{eq}} \right)^{3/2} \\
\end{equation}
with a high-$k$ cutoff at $k_\mathrm{hi} \approx 14 \mathcal{H}_\mathrm{eq} / \As^{1/4}$ associated with the light-crossing time of the largest virialized halos and a low-$k$ cutoff at $k_\mathrm{low} \approx 15 \mathcal{H}_\mathrm{eq}$, which is of order the collapse time of the latest-collapsing halos. Fitting to $\mathtt{S_{mid}}$ realizes an identical power law index with a $\sim10\%$ smaller amplitude, consistent with the higher resolution simulation capturing GW production from smaller halos forming at late times. In Fig.~\ref{fig:AsDep_observability} right, we compare this spectrum to projected observational sensitivities for several different values of $T_\mathrm{eq}$ and $\As$: depending on the frequency range, we find the SGWB from nonlinear EMDEs is potentially within reach for values of $\As\gtrsim 10^{-9}$. 
In the Supplementary Material, we discuss how the dominant role of the largest, latest-collaping structures renders this GW signal insensitive to deviations from scale invariance on much smaller scales. 

{\bf Discussion.}---%
Our results provide a conservative and model-independent lower bound on the SGWB 
resulting from structure formation during an EMDE.
The SGWB spectrum that we find reproduces the perturbative prediction on linear scales together with a larger contribution on nonlinear scales, which is dominated by the collapse of the largest and latest-forming halos.  We thus expect that, for scale-invariant spectra, the amplitude and location of the nonlinear peak depend on the properties of the EMDE only through $T_\mathrm{eq}$ and $k_\mathrm{NL}$, the largest scale to go nonlinear prior to reheating.  Even in our minimal scenario, we find that SGWBs from early structure formation may be within observational reach even for primordial curvature perturbations as small as those probed in the CMB, potentially opening a new window onto the expansion history of the universe prior to BBN. Intriguingly, a large $A_s \sim 10^{-3}$ initial curvature perturbation spectrum with a low reheat temperature $T_\mathrm{eq} \sim 10\,\mathrm{MeV}$ produces a SGWB with amplitude and frequencies broadly similar to the claimed detection by current pulsar arrays \cite{NANOGrav:2020spf, EPTA:2021crs, Goncharov:2021oub, Antoniadis:2022pcn}. However, a more complete characterization of the GW spectrum is required before it is possible to assess the goodness of fit subject to consistency both with primordial black hole constraints \cite{Josan:2009qn} and big bang nucleosynthesis \cite{deSalas:2015glj}.

In order to reliably predict the GW spectrum on scales smaller than the light-crossing time of the largest halos, it is necessary to go beyond the framework utilized here and perform a fully relativistic simulation of halo formation, ({\it e.g.}), through the approach of \cite{Adamek:2016zes}. 
A more relativistic treatment 
would also be necessary to make accurate predictions in scenarios of observational interest where the DM decays on a faster-than-Hubble timescale.
As a side benefit, such an approach would also correctly handle the difference between $\Phi$ and $\Psi$ that is generated by anisotropic stress during the process of collapse, although this effect is subleading.

Finally, we find that as long as the DR can be modeled as a perfect fluid, radiation sourced by the decay of collapsed structures does not provide an important source of GWs.
Because anisotropic stress is a leading source for tensor modes, our treatment of the DR as a perfect fluid at equilibrium at all times is a conservative assumption, but one that should be revisited in the future.

{\bf Acknowledgments.}---%
We particularly thank Jeppe Dakin for guidance regarding working with \CONCEPT and feedback on the first version of our manuscript. We thank Malte Buschmann, Patrick Draper, Alan Guth, Misha Ivanov, Toby Opferkuch, Evangelos Sfakianakis, Victoria Tiki, and Helvi Witek for useful conversations. N.F is supported by the U.S.~Department of Energy under grant DE-SC0010008. J.W.F was supported by a Pappalardo Fellowship. The work of B.L was supported in part by the U.S.~Department of Energy under Grant Number DE-SC0011640. The work of JS was supported in part by DOE grants DE-SC0023365 and DE-SC0015655. This work used NCSA Delta CPU at UIUC through allocation PHY230051 from the Advanced Cyberinfrastructure Coordination Ecosystem: Services \& Support (ACCESS) program, which is supported by National Science Foundation grants 2138259, 2138286, 2138307, 2137603, and 2138296.  JS gratefully acknowledges MIT's generous hospitality during the performance of this work. This research used resources of the Lawrencium computational cluster provided by the IT Division at the Lawrence Berkeley National Laboratory, both operated under Contract No. DE-AC02-05CH11231.

\bibliography{main}% Produces the bibliography via BibTeX.

\clearpage\onecolumngrid

\begin{center}
  \textbf{\large Supplementary Material for Stochastic GWs from Early Structure Formation}\\[.2cm]
  \vspace{0.05in}
  {Nicolas Fernandez, Joshua W. Foster, Benjamin Lillard, and Jessie Shelton}
\end{center}

\twocolumngrid

%%%%%%%%%% Merge with supplemental materials %%%%%%%%%%
\setcounter{equation}{0}
\setcounter{figure}{0}
\setcounter{table}{0}
\setcounter{section}{0}
\setcounter{page}{1}
\makeatletter
\renewcommand{\theequation}{S\arabic{equation}}
\renewcommand{\thefigure}{S\arabic{figure}}
\renewcommand{\theHfigure}{S\arabic{figure}}%
\renewcommand{\thetable}{S\arabic{table}}

\onecolumngrid

This Supplementary Material contains supporting material for the main Letter. In Sec.~\ref{sec:ICs}, we review the calculations in cosmological perturbation theory used to generate initial conditions for our $N$-body simulations with \CONCEPT. In Sec.~\ref{sec:NBody}, we describe our modifications to \CONCEPT~to enable decay radiation to be inhomogeneously sourced following the spatial distribution of decaying matter. In Sec.~\ref{sec:GWs}, we describe the equations of motion and algorithmic implementation of a lattice-based GW calculation in our modified version of \CONCEPT. In Sec.~\ref{sec:Soften}, we perform a systematic test of our simulation framework, demonstrating that the frequency spectrum of our induced gravitational waves is robust to our choice of force-softening in the \textit{N}-body dynamics. Finally, in Sec.~\ref{sec:AsScaling} we give an argument explaining the observed scaling of the energy density in gravitational waves with $\As$; we also determine the expected cutoff of our GW signal at the light-crossing time of collapsed halos.

\section{Gravitational Initial Conditions}
\label{sec:ICs}
In this section, we review the details of linear cosmological perturbation theory used to set the initial conditions for our simulations. Sec.~\ref{sec:Perturb} provides the defining equations of motion for perturbations entering the horizon in an EMDE, while Sec.~\ref{sec:Backscaling} reviews how initial conditions are determined for modes that are outside the horizon at the time our simulations begin.

\subsection{Linear Cosmology in the Conformal Newtonian Gauge}
\label{sec:Perturb}

We run our simulations in \CONCEPT\ in the conformal Newtonian gauge, 
\begin{equation}
\mathrm{d} s^2 = a^2 \left( -(1+2\Phi) \mathrm{d} \tau^2  + \left((1-2\Psi) \delta_{ij} + \frac{1}{2} h_{ij} \right) \mathrm{d}x^i \mathrm{d} x^j \right),
\end{equation}
and so we must review the details of linear perturbation theory in that gauge in order to properly set our initial conditions. We are interested in a cosmology with decaying dark matter and a perfect fluid of decay radiation, meaning that at in the linear theory the anisotropic stress $\sigma = 0$.  Then, using the conventions and definitions of Ref.~\cite{Ma:1995ey}, we have
\begin{gather}
    \Phi = \Psi \\
    \Phi' = -\mathcal{H} \left[\left(1 + \frac{k^2}{3 \mathcal{H}^2}\right) \Phi + \frac{1}{2} (\Omega_\mathrm{DM} \delta_\mathrm{DM} + \Omega_\mathrm{DR} \delta_\mathrm{DR})\right]
\end{gather}
where $\Phi$ and $\Psi$ are the scalar metric perturbations, $k$ the comoving wavenumber, $\mathcal{H}$ the conformal Hubble parameter, $\delta_i$ the density contrast in each species, and $\Omega_i$ the time-dependent fractional abundance of each species. The equations describing the evolution of perturbations in the decaying matter and decay radiation species in the conformal Newtonian gauge are given in \cite{Erickcek:2011us, Audren:2014bca, Poulin:2016nat}, which we follow closely. First, for the decaying matter we have
\begin{gather}
    \delta_\mathrm{DM}' = -\theta_\mathrm{DM} + 3 \Phi' - a \Gamma \Phi \\ 
    \theta_\mathrm{DM}' = - \mathcal{H} \theta_\mathrm{cdm} + k^2 \Phi
\end{gather}
where we have made the substitution $\Psi' = \Phi'$ in $\delta_\mathrm{DM}'$. Next we have
\begin{gather}
    \delta_\mathrm{DR}' = -\frac{4}{3} \left(\theta_\mathrm{DR} - 3 \Phi' \right) + a \Gamma \frac{\Omega_\mathrm{DM}}{\Omega_\mathrm{DR}} \left(\delta_\mathrm{DM} - \delta_\mathrm{DR} + \Phi\right) \\
    \theta_\mathrm{DR}' = k^2 \left( \frac{1}{4} \delta_\mathrm{DR} + \Phi \right) - a \Gamma \frac{\Omega_\mathrm{DM}}{\Omega_\mathrm{DR}} \left(\theta_\mathrm{DR} - \frac{3}{4} \theta_\mathrm{DM}\right).
\end{gather}
for the decay radiation.

The relevant initial conditions for our perturbations are provided in \cite{Barenboim:2021swl} for their scenario of ``Long Early Matter Domination", in which relevant modes enter the horizon during the EMDE. Here we summarize the key results. First, the initial potential perturbations $\Phi$ are related to the curvature perturbations $\mathcal{R}$ by 
\begin{equation}
    \Phi = \frac{3}{5} \mathcal{R}.
\end{equation}
Note that the prefactor of $3/5$ differs from the standard prefactor of $2/3$ relevant for modes that enter the horizon during radiation-domination. Next, at zeroth order in $\tau$, we have
\begin{gather}
    \delta_\mathrm{DM} = 2 \delta_\mathrm{DR} = -2 \Phi \\
    \theta_\mathrm{DM} = \theta_\mathrm{DR} = 0.
\end{gather}
In principle, we could evaluate the $\tau$-dependent terms that enter these initial conditions, but in practice, it suffices to evaluate our linear calculation starting at vanishingly small $\tau$ such that they are negligible.

\subsection{Growth Factors and Backscaling for Simulation Initial Conditions}
\label{sec:Backscaling}
We must also calculate the Newtonian linear growth factors for \CONCEPT\ in order for initial conditions to be generated from our perturbation spectrum. 

At first order, the growth factor $D_1(a)$ is given through 
\begin{equation}
    D_1'' + \mathcal{H}D_1' - 4 \pi G a^2 \bar\rho_\mathrm{DM} D_1 = 0.
\end{equation}
By convention, $D_1$ is normalized such that $D_1 = 1$ at $a = 1$ \cite{MBWgalaxy}. This defines one of the four boundary conditions. To fully specify the system, we take $D_1' = 0$ at $a = 0$. We also compute the associated growth factor $f_i$ defined by 
\begin{equation}
    f_1 = \frac{d \ln|D_1| }{d\ln a}
\end{equation}
following the \texttt{CLASS} conventions.

As is standard for cosmological $N$-body simulations, our initial conditions are generated using the Zel'dovich approximation using \CONCEPT's built-in routines. However, this presents a challenge for our simulations, which attempt to resolve both small-scale modes that experience late-time nonlinearities and the horizon scale at the time of reheating. While keeping $\Delta_k^2 \ll 1$ on small-scales requires us to begin our simulations at small $a_i$, our simulation volume will then contain many initially superhorizon modes that experience unphysical linear growth when evolved under Newtonian gravitational dynamics.

\begin{figure*}[!ht]
\includegraphics[width=.49\textwidth]{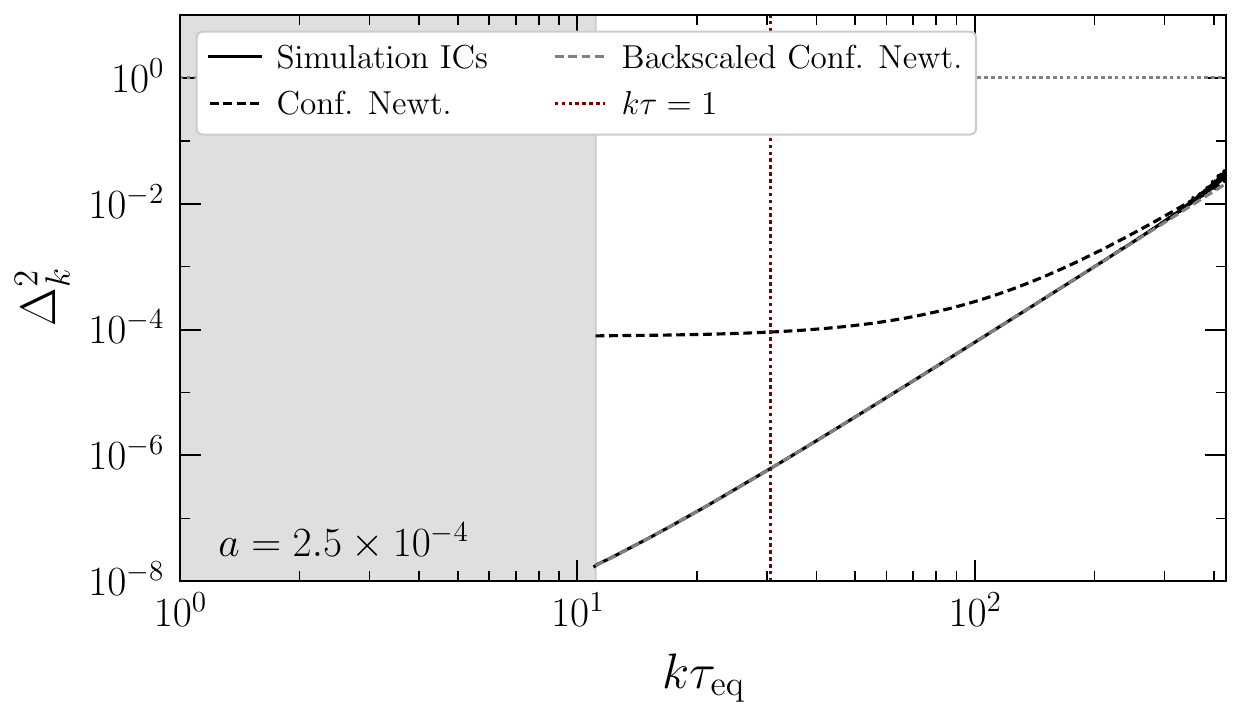}
\includegraphics[width=.49\textwidth]{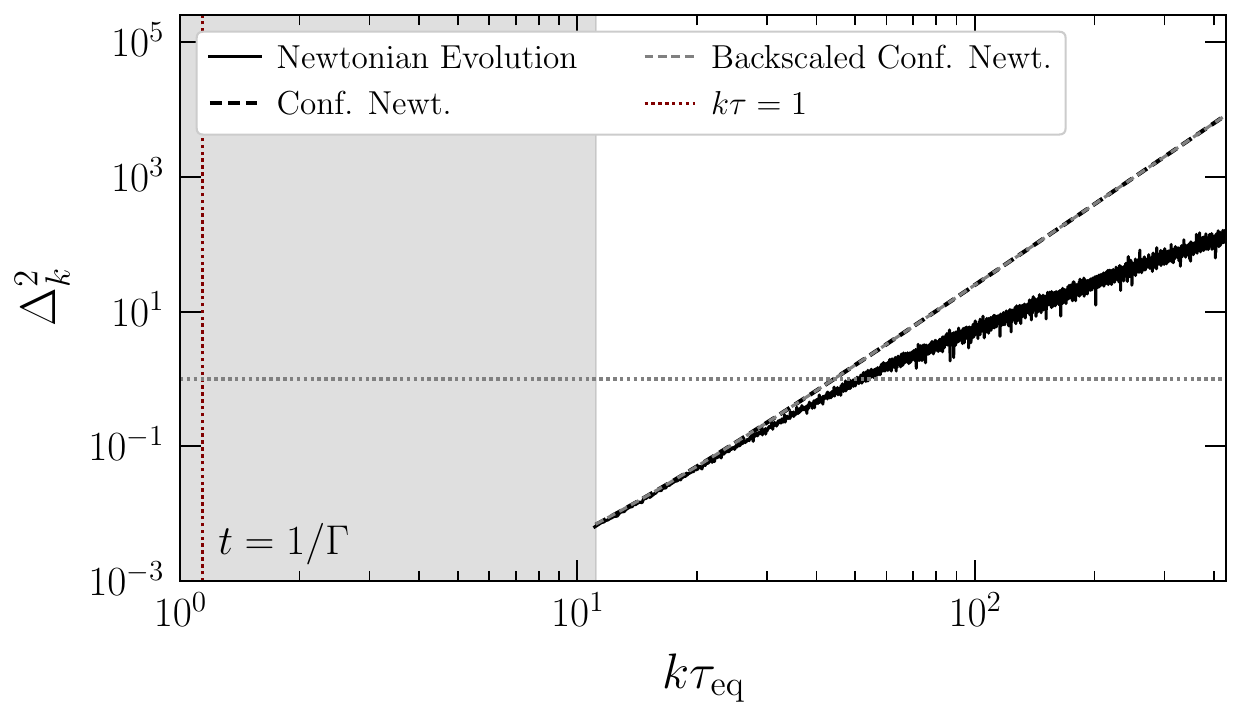}
\caption{(\textit{Left.}) The initial conditions of a simulation starting in an EMDE at $a = 2.5 \times 10^{-4}$ compared to the conformal Newtonian predictions for the matter power spectrum. Rather than generating the initial conditions from this prediction, we generate them according to a back-scaled matter power spectrum engineered to achieve the correct matter power spectrum at the decay time of metastable matter. As a reference, the horizon scale is indicated by a dotted red line. (\textit{Right.}) The matter power spectrum realized by the simulation at the decay time as compared to the conformal Newtonian prediction and the prediction of the back-scaled initial matter power spectrum advanced under Newtonian dynamics. These predictions agree at the subpercent level and match the matter power spectrum realized in the simulation on linear scales. As before, the horizon scale at this time is indicated by a dotted red line, and it is apparent that all modes are safely subhorizon and hence will continue to evolve, to good approximation, subject to Newtonian dynamics. In our simulations, we do not include modes at $k$ smaller than $10/\tau_{\mathrm{eq}}$, indicated by the shaded region in the plots.}
\label{fig:PowerSpectra}
\end{figure*}

To address this challenge, we employ the standard ``back-scaling" trick, in which our initial conditions are set such that the correct linear matter power spectrum at matter-radiation equality is achieved when modes are evolved under Newtonian dynamics \cite{Fidler:2017ebh}. This amounts to a scale-dependent suppression of the initial power of super-horizon modes so that they grow to their correct values at late times. An example of this is shown in Fig.~\ref{fig:PowerSpectra}. In the left panel, the initial conditions in an EMDE at $a = 2.5 \times 10^{-4}$ are set not according to the properly evaluated cosmological perturbation theory in the conformal Newtonian gauge but instead to the conformal Newtonian gauge initial conditions backscaled by dividing out by the Newtonian linear growth factor. As shown in the right panel, at the decay time, the power has grown under the Newtonian dynamics, achieving the predicted linear matter power spectrum on scales that remain linear. As is typical, power in the nonlinear regime is suppressed with respect to the linear theory prediction. In our simulations, we never provide power to modes with $k \tau_\mathrm{eq} < 10$, which ensures subpercent accuracy in the matter power spectrum realized at matter-radiation equality as compared to its expectation computed with linear cosmological perturbation theory. 

\section{N-Body Simulation with Decaying Matter and Decay Radiation in CONCEPT}
\label{sec:NBody}

In this section, we describe the procedure through which we use \CONCEPT\ to model the gravitational evolution of decaying matter and decay radiation. Throughout this section, we will work with \CONCEPT's fluid variables $\varrho$, $J^{i}$ and $\varpi$ \cite{Dakin:2017idt}, which are defined in the following way. First, for a fluid with a spatially-averaged physical density $\bar \rho$ that is some general function of scale factor $a$, the effective equation of state is defined as 
\begin{equation}
    \varpi(a) \equiv \frac{\ln \bar\rho(a=1) / \bar \rho(a)}{3 \ln a}-1.
\end{equation}
From this definition, the conserved energy density $\varrho$ is defined in terms of the physical density as 
\begin{equation}
    \varrho(a) \equiv a^{3(1 + \varpi(a))} \rho(a)
\end{equation}
where we have dropped the overbar as this definition extends to the spatially inhomogeneous density field. Finally, the conserved current $J^i$ is defined in terms of the velocity perturbation $v^{i}$ by
\begin{equation}
    J^{i}(a) = a^{4} (\rho + P) v^{i}
\end{equation}
where $\rho$ is the physical density and $P$ the pressure.

\subsection{Real-time Decay Sources}
We begin our simulations using \CONCEPT's built-in initial condition generator at cosmic times well before the decaying dark matter lifetime $1/\Gamma$. Supposing that at time $t_1$ the physical radiation density is $\rho_1$, then we have
\begin{equation}
\varrho^\mathrm{DR}_1 = a^{3(1+\varpi_1)} \rho^\mathrm{DR}_1
\end{equation}
where $\varpi_1$ is the effective equation of state for decay radiation at time $t_1$. Now consider the radiation density at time $t_2$. Supposing no radiation were sourced, then we would have
\begin{equation}
    \rho^\mathrm{DR}_2 = \left( \frac{a_1}{a_2} \right)^{3 (\omega+1)} \rho^\mathrm{DR}_1 = \left( \frac{a_1}{a_2} \right)^{3 (\omega+1)} a^{-3(1 + \varpi_1)} \varrho^\mathrm{DR}_1.
\end{equation}
We can then calculate the contribution of the radiation present at $t_1$ to the abundance of radiation $\varrho_2$ at time $t_2$ by
\begin{equation}
    \varrho^\mathrm{DR}_2 = \frac{a_1^{3(\omega - \varpi_1)}}{a_2^{3(\omega - \varpi_2)}} \varrho^\mathrm{DR}_1.
\end{equation}
Hence, by applying the rescaling factor defined by 
\begin{equation}
    R(a_1, a_2) \equiv \frac{a_1^{3(\omega - \varpi_1)}}{a_2^{3(\omega - \varpi_2)}}
\end{equation}
to $\varrho^\mathrm{DR}_1$, we calculate the contribution to $\varrho^\mathrm{DR}_2$ associated the radiation present at time $t_1$ appropriately redshifted to time $t_2$. Now we must calculate the contribution of decaying dark matter to the decay radiation $\varrho_2$ over the interval $(t_1, t_2)$. Recognizing that the energy density must be preserved, we can immediately identify that the spatially averaged contribution of the decaying dark matter must be given by
\begin{equation}
    \bar \varrho_\mathrm{inj.} = [1-R(a_1, a_2)]  \bar \varrho^\mathrm{DR}_1.
\end{equation}
Then we can write down the update step from spatially inhomogeneous $\varrho_1$ to $\varrho_2$ by
\begin{equation}
    \varrho_2(x) = R(a_1, a_2) \varrho_1(x) + \frac{[1-R(a_1, a_2)] \bar \varrho_1}{\bar \varrho_1^\mathrm{DM}} \varrho_1^\mathrm{DM}(x),
\end{equation}
which explicitly preserves $\varrho$ as time-independent as is expected while appropriately balancing redshifting existing radiation and the injection of new radiation. 

The momentum densities are more transparently sourced. For the decaying dark matter, the particle mass is given by
\begin{equation}
    m(a) = m(a = 1) a^{-3 \varpi(a)}
\end{equation}
where $\varpi$ now denotes the decaying dark matter equation of state. Then the change in the momentum from $t_1$ to $t_2$ is given by
\begin{equation}
    \Delta J^{i}_\mathrm{DM} = J^{i}_\mathrm{DM}(t_1) \frac{a_1^{3 \varpi(a_1)}}{a_2^{3 \varpi(a_2)}}.
\end{equation}
Hence, we directly add
\begin{equation}
    J^{i}_\mathrm{DR}(t_2) = J^{i}_\mathrm{DR}(t_1) + J^{i}_\mathrm{DM}(t_1) \left(1-\frac{a_1^{3 \varpi(a_1)}}{a_2^{3 \varpi(a_2)}}\right)
\end{equation}
to impose momentum conservation.

\subsection{Time-Stepping Scheme for N-Body Simulation}
In order to properly source the decay radiation, the particle positions and momenta must be meshed field data living on the same lattice as the decay radiation fluid. For an accurate calculation of the source terms, the particle positions and momenta must be evaluated simultaneously, but this is generically not true in \CONCEPT's default leapfrog integration scheme. To address this, we modify \CONCEPT~to use a synchronized Kick-Drift-Kick (KDK) scheme. In this scheme, after each KDK, the positions and momenta are synchronized, allowing for meshing to be performed prior to the first kick of the next integration step while preserving the symplectic property of the integrator.

In actuality, the decay injection should appear in the equations of motion for the decay radiation fluid, but accomplishing this within the KDK framework is not possible. We therefore perform a second order Strang splitting by sourcing radiation over an interval $\Delta t/2$, followed by a KDK step of size $\Delta t$, and then sourcing radiation again over an interval $\Delta t/ 2$. An additional advantage of this approach is that it allows us to leave the internal fluid evolution of \CONCEPT~unmodified and make use of the total variation-diminishing Kurganov-Tadmor method with flux limiters without modification.

\section{Gravitational Wave Calculation from N-Body Simulation}
\label{sec:GWs}

In this section, we describe the how the GW energy density is calculated using states of the $N$-body simulation performed using \CONCEPT. 

\subsection{Gravitational Wave Equations of Motion}
From \cite{Baumann:2007zm}, we may obtain the equations of motion for the gravitational waves. First, we have the perturbed Einstein tensor at second order:
\begin{equation}
    G_{ij} = \frac{1}{a^2}\left[\frac{1}{4}\left(h_{ij}'' + 2 \mathcal{H} h_{ij}' - \nabla^2 h_{ij}\right) + 4 \Phi \partial_i \partial_j \Phi + 2 \partial_i \Phi \partial_j \Phi \right].
\end{equation}
Relating this to the transverse-traceless component of the stress-energy tensor, we have
\begin{equation}
    h_{ij}'' + 2 \mathcal{H} h_{ij}' - \nabla^2 h_{ij} = \left[32 \pi G a^2 T_{ij} - 16 \Phi \partial_i \partial_j \Phi - 8 \partial_i \Phi \partial_j \Phi\right]^{TT},
\end{equation}
where the $TT$ superscript represents the transverse-traceless part. In order to use the trick of linearity of the equations of motion with respect to the transverse-traceless projector, as developed in \cite{Garcia-Bellido:2007fiu, Dufaux:2007pt}, we define the projection operators 
\begin{gather}
    \tilde{\mathcal{S}}_{ij}^{TT}(\mathbf{k}) =\Lambda_{ij,lm}(\mathbf{k})  \tilde{\mathcal{S}}_{lm}(\mathbf{k}) \\
    \Lambda_{ij,lm}(\mathbf{k}) = P_{il}(\mathbf{k})P_{jm}(\mathbf{k}) - \frac{1}{2}P_{ij}(\mathbf{k}) P_{lm}(\mathbf{k})\\
    P_{ij}(\mathbf{k})= \delta_{ij} - \hat{\mathbf{k}}_i \hat{\mathbf{k}}_j.
\end{gather}
Defining an auxiliary field $u_{ij}$ where $u_{ij}$ is defined by
\begin{equation}
    h_{ij} = \Lambda_{ij, lm} u_{lm},
\end{equation}
we can then evolve the equation of motion 
\begin{equation}
    u_{ij}'' + 2 \mathcal{H} u_{ij}' - \nabla^2 u_{ij} = 32 \pi G a^2 T_{ij} - 16 \Phi \partial_i \partial_j \Phi - 8 \partial_i \Phi \partial_j \Phi.
\end{equation}
using the full source tensor to source the $u_{ij}$. A final convenience is enabled by defining a conformally-rescaled $\phi_{ij} = a u_{ij}$ \cite{Figueroa:2020rrl}. Our equations of motion then take the form
\begin{equation}
    \phi_{ij}'' = \nabla^2 \phi_{ij} + \frac{a''}{a}\phi_{ij}  +  32 \pi G a^3 T_{ij} - 8 a \bigg(\partial_i \Phi \partial_j \Phi + 2  \Phi \partial_i \partial_j \Phi\bigg).
\end{equation}
It now remains to evaluate the contributions to the right-hand side of this equation.

\subsection{Stress-Energy of Particle Species}
We start with \cite{Weinberg:1972kfs}, where the stress-energy tensor for a point-particle is given by
\begin{equation}
T_{ij}(\vec{r}, t) = \frac{m}{\sqrt{-g}} \frac{dx_i}{dt}\frac{dx_j}{dt} \frac{dt}{d\sqrt{-s^2}} \delta^{(3)}(\vec{r}-\vec{x})
\end{equation}
It is important that we carefully consider the structure of the particle data that \CONCEPT\ evaluates, which are the positions of particles in comoving coordinates, $\mathbf{x}$ and the associated conjugate momentum $\mathbf{q} = a^2 m \dot{\mathbf{x}}$ where the dot indicates differentiation with respect to cosmic time. We can then identify this particle location as $\mathbf{x}^{i}$ and the momentum as $\mathbf{q}_{i} = m \dot{\mathbf{x}}_i$ when we use the line element
\begin{equation}
    ds^2 = -dt^2 + a^2 d\mathbf{x}^2.
\end{equation}
Using this coordinate system, we can substitute into the definition of the stress-energy tensor to obtain
\begin{equation}
T_{ij}(\vec{r}, t) = \frac{1}{a^3}\frac{q_i q_j }{m} \frac{dt}{d\sqrt{-s^2}} \delta^{(3)}(\vec{r}-\vec{x}).
\end{equation}
Making a nonrelativistic approximation, we have
\begin{equation}
T_{ij}(\vec{r}, t) \approx \frac{1}{a^3}\frac{q_i q_j }{m} \delta^{(3)}(\vec{r}-\vec{x}).
\end{equation}
We must also remember that $m$ varies with redshift to describe decaying dark matter so $m \rightarrow m(a)$. We are also now free to change from cosmic time to conformal time without any change in the spatial components of $T_{ij}$. We then have
\begin{equation}
T_{ij}(\vec{r}) \approx \frac{1}{a^3}\frac{q_i q_j }{m(a)} \delta^{(3)}(\vec{r}-\vec{x}).
\end{equation}
where we have suppressed the temporal coordinate. It is then straightforward to promote this to a sum over particles as
\begin{equation}
T_{ij}(\vec{r}) = \frac{1}{a^3 m(a)} \sum_{n} q^{(n)}_i q^{(n)}_j \delta^{(3)}(\vec{r}-\vec{x}^{(n)}).
\end{equation}
where the superscript $n$ indexes the particles. In \CONCEPT, the time-dependent mass is given by
\begin{equation}
    m(a) = m_0 a^{-3 \varpi(a)},
\end{equation}
where $m_0$ is the particle mass at time $t = 0$. We can then substitute this in to obtain 
\begin{equation}
T_{ij}(\vec{r}) = \frac{a^{3(\varpi - 1)}}{m_0} \sum_{n} q^{(n)}_i q^{(n)}_j \delta^{(3)}(\vec{r}-\vec{x}^{(n)}).
\end{equation}
Finally, in the \textit{N}-body scheme, we replace the delta function with a window function in our meshing procedure.

\subsection{Stress-Energy of Fluid Species}

The stress-energy tensor for a perfect fluid is given by
\begin{equation}
    T^{ij} = (1 + \omega) \rho U^i U^j + \omega \rho g^{ij},
\end{equation}
where $\omega$ is the equation of state parameter of the fluid, $\rho$ is its density, $U$ its four-velocity and $p=\omega \rho$ \cite{2013rehy.book.....R}. We neglect the $\omega \rho g^{ij}$ term as its contributions to anisotropic sources enter at a higher order than the accuracy to which we work. The spatial components of the fluid four-vector in a perturbed universe are given by
\begin{equation}
    U^{i} = \frac{1}{a} u^{i} ,
\end{equation}
where $u^{i}$ is the velocity perturbation \cite{Baumann:2022mni}. Substituting this in, and lowering indices by multiplying by a factor of $a^4$, we have
\begin{equation}
T_{ij} = a^2 (1+\omega) \rho u^i u^j.
\end{equation}
Finally, substituting in the internal \CONCEPT\, definitions for $\varrho$ and $J$ \cite{Dakin:2017idt} that are provided in Sec.~\ref{sec:NBody}, we obtain
\begin{equation}
    T_{ij} = a^{3 (\varpi - 1)} \frac{J^{i} J^{j}}{(1+\omega) \varrho},
\end{equation}
where $\varpi$ is the effective equation of state for the fluid species.

\subsection{Calculating Bardeen Potentials in the Newtonian Approximation}

In the Newtonian approximation for the Bardeen potential, we merely solve
\begin{equation}
    \nabla^2 \Phi = 4 \pi G a^2 (\rho_\mathrm{DM} + \rho_\mathrm{DR}).
\end{equation}
Since the quantities $\rho_\mathrm{DM}$ and $\rho_\mathrm{DR}$ are evaluated in real time by our $N$-body simulation, so too can be $\Phi$. In principle, we could take advantage of \CONCEPT\'s internal calculation as part of its gravitational evolution, but the recommended treatment is to take the matter potential to be twice as spatially well-resolved as compared to the radiation potential and treat the two separately \cite{Dakin:2021ivb}. Since all of our potentials for calculating GWs must live on identically-sized grids, we choose then to simply perform our own evaluation. 

\subsection{Validity of the Second-Order Source Tensor}
In our simulation, we evaluate the gravitational waves using the equations of motion
\begin{equation}
    h_{ij}'' + 2 \mathcal{H} h_{ij}' - \nabla^2 h_{ij} = 4 \mathcal{S}_{ij}^\mathrm{TT} ,
    \label{Eq:EoM}
\end{equation}
with the source tensor 
\begin{equation}
    \mathcal{S}_{ij} = 8 \pi G a^2 \mathcal{T}_{ij} -4 \Phi \partial_i \partial_j \Phi -2  \partial_i \Phi\partial_j \Phi,
    \label{eq:SourceTensor}
\end{equation}
Our simulations will begin in the regime of linear cosmology with small $\delta$, but $\delta$ will grow such that, on some scales, $\delta > 1$, and the gravitational dynamics of the matter distribution becomes nonlinear. As we now argue, the form of our source tensor is consistent with weak gravitational field expansions in both the large and small matter overdensity regimes.

In the limit of small matter overdensities $\delta$, the standard cosmological perturbation theory approach is to expand in $\delta$ \cite{Malik:2008im}. Perfect fluids do not support tensor modes at first order in $\delta$, and thus the leading contribution of perfect fluids to the source tensor is second order, 
%and the second-order terms in the stress-energy tensor 
$\mathcal{T}_{ij} \sim \rho u_i u_j \approx \bar \rho u_i u_j$, where $u$ is the bulk velocity. 
Also at second order in this expansion scheme, quadratic combinations of first-order perturbations in $\Phi$ act as second-order sources for the tensor perturbations \cite{Ananda:2006af, Baumann:2007zm,Malik:2008im}. For large $\delta$, standard cosmological perturbation theory breaks down, but an alternative velocity expansion appropriate for this regime in the conformal Newtonian gauge is developed in \cite{Baumann:2010tm}. Within virialized structures, $\Phi \propto u^2$; meanwhile the Poisson equation indicates each spatial derivative carries order $1/u$. Hence, $\mathcal{T}_{ij}$ remains second-order, though now in $u$, as do the terms that are quadratic in spatial derivatives of $\Phi$. We conclude that the source tensor given in Eq.~\ref{eq:SourceTensor} fully captures the leading order tensor mode production in both limits of $\delta$ we consider in this work.

\subsection{Numerical Integrators for Gravitational Waves}
Second-order partial differential equations that do not include a dissipative term are known as ``special" differential equations, and there exists a class of methods specifically designed for their numerical integration. A standard choice would be to use a Runge-Kutta-Nystrom method, but this would require evaluated intermediate steps of our $N$-body simulation in order to accurately evaluate the source terms appearing in the different integration stages. This is both challenging and computationally impractical. 

We instead elect to use a St\"ormer-Cowell method, which is a linear multistep method that can be directly employed subject to the restriction that the time-stepping remains uniform over the duration of the simulation. However, this does not introduce appreciable additional cost relative to a simulation that already resolves the decay radiation. We find that a Courant-Friedrichs-Lewy (CFL) condition of 
\begin{equation}
    \frac{\Delta x}{\Delta \tau} < C \equiv .05,
\end{equation}
provides good stability for our numerical integration, while an almost identical CFL condition is implemented for nonlinear realizations of fluids in \CONCEPT. That the requisite timesteps are similar is unsurprising as both conditions are designed to resolve the time evolution of relativistic dynamics. For additional stability, we perform our integration as a Predict-Evaluate-Correct-Evaluate method using a $5^\mathrm{th}$ order forward step followed by a  $6^\mathrm{th}$ order backward step.

\section{Dependence on Force Softening Length}
\label{sec:Soften}

\begin{figure*}[!ht]
\includegraphics[width=0.8\textwidth]{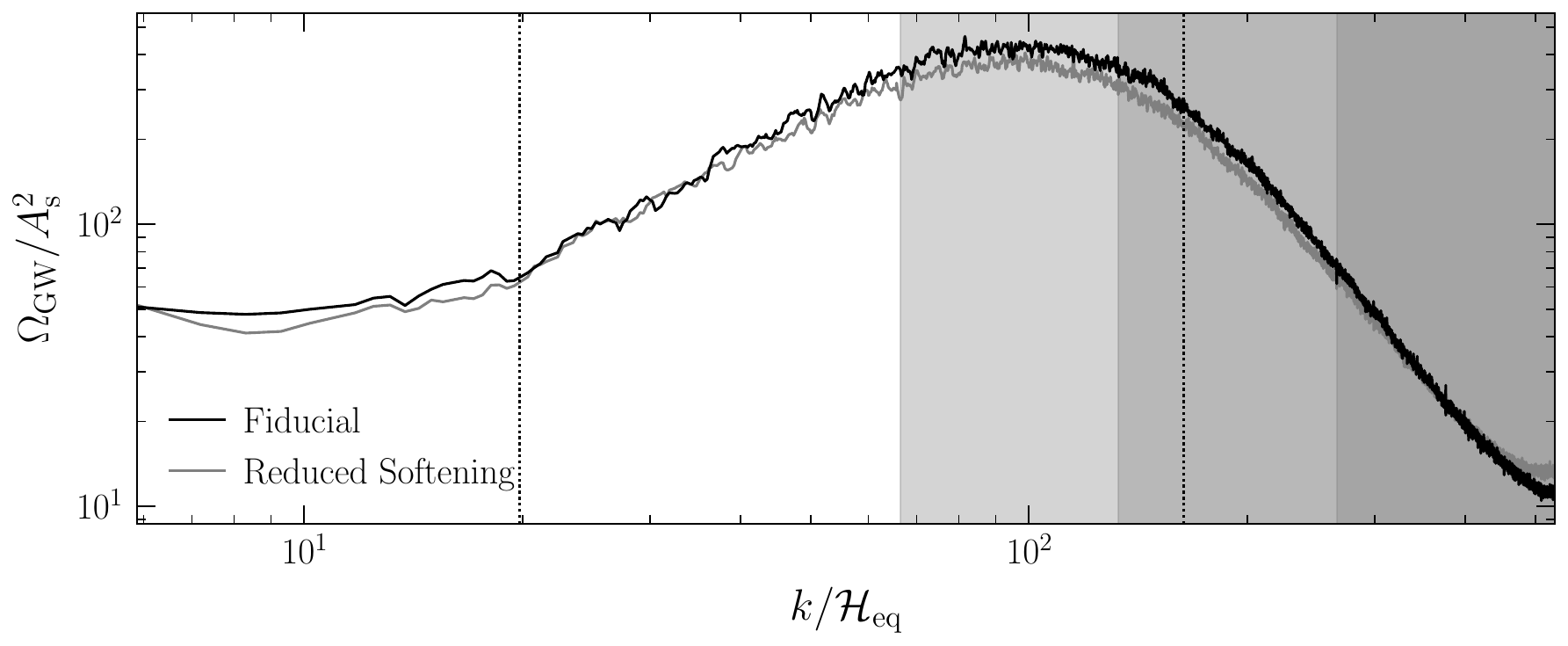}
\caption{A comparison of the gravitational wave spectrum for our fiducial simulation $\mathtt{S_{mid}}$ with an otherwise identical simulation using a force-softening length shrunk by a factor of 2. For clarity, the spectra have been smoothed with a Savitzky-Golay filter. The dotted vertical lines indicate (left) the value of $k_\mathrm{NL}$ and the light-crossing cutoff $\mathrm{k}_\mathrm{cross}$. The light grey, medium grey, and dark grey bands indicate the range of $k$ which are a factor of 8, 4, and 2 from the Nyquist frequency, respectively. Good agreement between the induced gravitational wave spectra is observed.}
\label{fig:ForceSoftening}
\end{figure*}

As a systematic test, we perform an identical simulation to that of $\mathtt{S_\mathrm{mid}}$ but we shrink the force-softening length used in the $P^{3}M$ method from its default value in \CONCEPT$\,$ of $0.025 L / N^{1/3}$ to $0.0125 L / N^{1/3}$. The force-softening length, which sets the scale of energy-nonconservation, is well below our lattice resolution and even further below the nonlinear scale, and so we do not anticipate it has a strong effect on our results. This is directly observed in Fig.~\ref{fig:ForceSoftening}. 

\section{Parametric Dependence of the Collapse Signal}
\label{sec:AsScaling}

The observed $\As$-dependence of the GW emission from collapsing halos can be understood using a straightforward Press-Schechter argument~\cite{Press:1973iz}. 

We start by estimating the properties of the typical halos collapsing at redshift $a<\arh$. 
Consider the variance of the density field smoothed on a comoving scale $R$ at scale factor $a$:
\beq
\label{eq:variance}
\sigma^2 (a,k) = \int \frac{dk^3}{(2\pi)^3} W(kR) \langle \delta (a,k) ^2 \rangle,
\eeq
where $\delta(a,k)$ is the matter density perturbation, and we take the window function $W(kR)$ to be a Fourier-space top hat for simplicity, with upper limit $k_R\equiv 1/R$.  For long EMDEs, we can express  $\delta(a,k)$ in terms of the primordial curvature perturbation $\mathcal{R}_k$  as
\beq
\delta(k,a) = \frac{6}{5} \frac{a}{a_{hor}} \mathcal{R}_k = \frac{6}{5} \frac{a}{\arh}\frac{k^2}{\mathcal{H}_\mathrm{eq}^2} \mathcal{R}_k.
\eeq
Using this expression for $\delta$ in Eq.~\ref{eq:variance} for the variance and plugging in the primordial curvature power spectrum then gives
\beq
\sigma^2 (a,k_R) = \int_0^{k_R} \frac{ dk}{k} \left[  \frac{36}{25} \left(\frac{a}{\arh}\right)^2 \frac{k^4}{\mathcal{H}_\mathrm{eq}^4}  \right] \As,
\eeq
where we have taken the amplitude of the primordial power spectrum, $\As$, to be constant over the range of $k$ of interest. 
Doing the $k$-integral then gives simply
\beq
\label{eq:sigma}
\sigma (a,k_R) = \frac{3}{5} \left(\frac{a}{\arh}\right) \frac{\sqrt{\As}}{\mathcal{H}_\mathrm{eq}^2} k_R^2.
\eeq
The scale $R_*(a)$ for typical halos collapsing at the scale factor $a$ follows by setting this variance equal to the critical threshold $\delta_c = 1.686$~\cite{MBWgalaxy}:
\beq
\label{eq:Rstar}
R_*(a) = \left[ \frac{3}{5\delta_c}\left(\frac{a}{\arh}\right) \frac{\sqrt{\As}}{\mathcal{H}_\mathrm{eq}^2} \right]^{1/2}.
\eeq
Consistent with our choice of the fourier top-hat window function, these structures have mass $M_*(a) = 6\pi^2 \bar \rho (a) (a R_*)^3 $, where $\bar \rho (a)$ is the homogeneous background matter density. In terms of reheating-era quantities, this mass is 
\beq
\label{eq:mstar}
M_* (a) = \frac{9 \pi a_{\mathrm{eq}} \mathcal{H}_{\mathrm{eq}}^2}{4 G}  \left[ \frac{3}{5\delta_c}\left(\frac{a}{\arh}\right) \frac{\sqrt{\As}}{\mathcal{H}_\mathrm{eq}^2 }\right]^{3/2}.
\eeq
The comoving number of halos in a given mass interval is given by the Press-Schechter mass function,
\beq
\frac{dn}{dM} = -\sqrt{\frac{2}{\pi}} \frac{\rho_m a^3}{M} \frac{\delta_c}{\sigma^2} \frac{\partial \sigma}{\partial M} e^{-\delta_c^2/2\sigma^2} .
\eeq
We will be interested in this quantity for $a\approx \arh$. For collapsed structures, the exponential is an $\mathcal{O}(1)$ factor. Then, we estimate the comoving number density of halos of mass $M$ collapsing near $\arh$ as:
\beq
\label{eq:dndm}
\left.\frac{dn}{dM}\right|_{\arh} \approx \frac{2}{3} \sqrt{\frac{2}{\pi}} \frac{3 a_{\mathrm{eq}}\mathcal{H}_{\mathrm{eq}}^2}{8\pi G M^2} \frac{\delta_c}{\sigma (\arh,M)} .
\eeq
Sizeable anisotropic stress is generated in the process of shell crossing during a halo's initial collapse at scale factor $a_c$. For a halo of mass $M$ and (physical) radius $\tilde R$, this anisotropic stress can be parameterized by $ \epsilon M^2/\tilde R$, i.e., an undetermined efficiency factor $\epsilon$ times the energy of the gravitationally-bound system.  We estimate that this anisotropic stress remains sizable over the collapse timescale $t_c (a_c)$, and thus actively sources gravitational waves over this timescale.  Then we can approximate the outgoing gravitational wave a radial distance $x\gg\tilde R$ from the halo as  
\beq
h (x) \sim \frac{G}{x} \left(\frac{\epsilon G M^2}{\tilde R}\right).
\eeq
The gravitational wave luminosity $\mathcal{L}$  then follows as
\beq
\label{eq:lumi}
\mathcal{L} (M) \sim\frac{x^2}{G}|\dot h|^2 \sim \frac{\epsilon^2}{G} \left(\frac{d}{dt} \frac{G^2 M^2}{\tilde R} \right)^2 \sim \epsilon^2 G \left(\frac{G M^2}{\tilde R t_c}  \right)^2.
\eeq
This luminosity scales rapidly with increasing halo mass. The gravitational waves from halo collapse will therefore be dominated by the largest structures, which form at the latest possible times, i.e., around the reheating timescale $t_\phi= 1/{\Hrh}$. %
Since the signal is thus dominated by emission from structures that form immediately before the return to radiation domination, we can neglect the dilution of gravitational waves through red-shifting, and estimate the energy density in gravitational waves as
\begin{align}
\Delta \rho_{\mathrm{GW}} &\approx  \epsilon^2 \frac{1}{\arh^3} \int_0^{M_* (\arh)} dM \frac{dn}{dM} \mathcal{L}(M) t_\phi.
\end{align}
Here in using Eq.~\ref{eq:lumi} for the halo luminosity $\mathcal{L}(M)$, the physical radius $\tilde R$ is given by $\tilde R = a_{c} R_* (a_c)$. This integral is dominated by its upper endpoint, giving the parametric dependence
\begin{align}
\label{eq:drho}
\Delta \rho_{\mathrm{GW}}     
 &\sim  \frac{1}{t_\phi} \rho(\arh) \frac{H_{\mathrm{eq}}^2 G^3}{\sqrt{\As}} M^3_* (\arh),
\end{align}
or, using $t_\phi = 1/\Hrh$ and $M_* (\arh)\sim \As^{3/4}/(\Hrh G)$,
\beq
\Omega_{\mathrm{GW}} \propto  \As^{7/4} .
\eeq
This dependence on $\As$ precisely matches what we observe in our numerical data.

Beyond the $\As$ dependence, this argument also predicts that the dominant contribution to the energy density in GWs is determined by the fraction of energy bound up in collapsing gravitational structures near reheating, and is not directly dependent on the timescale of reheating itself. Thus while $H_\mathrm{eq}$ sets the overall frequency scale for the GW spectrum, the amplitude of the spectrum depends only on $\As$. Moreover, since the GW emission is dominated by the largest and latest-forming structures, it depends on the duration of the EMDE only through the comoving wavenumber $k_\mathrm{NL}$ of the typical structures collapsing at reheating.

From Eq.~\ref{eq:Rstar}, $k_\mathrm{NL}$ is related to $\As$ and the reheating timescale through
\beq
k_\mathrm{NL} \approx  1.7\times \frac{\mathcal{H}_\mathrm{eq}}{\As^{1/4}},
\eeq
and thus we can express the energy density in GWs in terms of the hierarchy between $H_\mathrm{eq}$ and $k_\mathrm{NL}$,
\beq
\Omega_{\mathrm{GW}} \propto \left(\frac{\mathcal{H}_\mathrm{eq}}{k_\mathrm{NL} }\right)^{7}.
\eeq
The mode $k_\mathrm{NL}$ entered the horizon during matter domination when $k_\mathrm{NL} = \mathcal{H}(a_\mathrm{NL})$ at scale factor $a_\mathrm{NL} \approx a_\mathrm{eq} \As^{1/2} / 3$.
Thus the minimum duration of an EMDE that forms nonlinear structure can be expressed as
\beq
\frac{a_\mathrm{eq}}{a_i} \gtrsim \frac{3}{\As^{1/2}},
\eeq
where $a_i$ is the scale factor at the onset of the EMDE.

The light-crossing time for a collapsing halo at $a = a_\mathrm{eq}$ associated with the mode $k_\mathrm{NL}$ can be determined from its virialized overdensity parameter relative to the matter fraction, given by 
\begin{equation}
    \Delta_c = \frac{18 \pi^2 + 60(\Omega_m-1) - 32 (\Omega_m-1)^2 }{\Omega_m}
\end{equation}
with $\Delta_c \approx 280$ at matter-radiation equality \cite{Bryan:1997dn}. Assuming spherical collapse, the nonlinear scale $k_\mathrm{NL}$ and the comoving virial radius $R_\mathrm{vir}$ are related by
\begin{equation}
    \frac{6 \pi^2 a_\mathrm{eq}^3 \bar\rho(a_\mathrm{eq})}{k_\mathrm{NL}^3} = \frac{4 \pi}{3} \Delta_c  a_\mathrm{eq}^3 \bar\rho(a_\mathrm{eq}) R_\mathrm{vir}^3
\end{equation}
so that the virial diameter is given by
\begin{equation}
    D_\mathrm{vir} \approx 0.4 \times \frac{\As^{1/4}}{\mathcal{H_\mathrm{eq}}}.
\end{equation}
The comoving wavenumber associated with gravitational waves induced by dynamics at the light-crossing time $t_\mathrm{cross} = D_\mathrm{vir}/c$ is then
\begin{equation}
     k_\mathrm{cross} = \frac{2 \pi}{D_\mathrm{vir}} \approx 14 \times \frac{\mathcal{H}_\mathrm{eq}}{\As^{1/4}}.
\end{equation}
Finally, it is worth pointing out that we have taken the asymmetry factor $\epsilon$ to be scale-invariant. This is likely to be a good assumption for a scale-invariant primordial spectrum but may need to be revisited for sharply-peaked curvature power spectra. However, we expect the leading effect of a scale-varying primordial power spectrum to be modifications to the redshift dependence of halo formation; scale variation in $\epsilon$ will arise as a consequence of this scale-dependent collapse history.

Since we find our GW signal is dominated by the largest and latest-collapsing halos, we can conclude that the signal is insensitive to deviations from scale invariance for wavenumbers $k\gg k_\mathrm{NL}$. In other words, provided that the primordial curvature power spectrum is nearly scale-invariant for $k\sim k_\mathrm{NL}$, the signal is insensitive to the details of any small-scale cutoff. This property ensures that the collapse GW signal applicable to a wide range of microphysical models that realize EMDEs.  Similarly, the GW signal is insensitive to the details of the transition into the EMDE, provided only that the duration of the EMDE is long enough that $k_\mathrm{NL}$ enters the horizon during the EMDE itself. 

The estimate of the GW emission from halo collapse made in Ref.~\cite{Jedamzik:2010hq} differs from that presented here primarily through their assumption that the emission from a collapsing halo can be associated with a single frequency $f \sim 1/(2\pi t_c)$.\footnote{The GW luminosity of a single collapsing halo estimated in Ref.~\cite{Jedamzik:2010hq} is parametrically the same as the one adopted here, since the characteristic halo-crossing timescale, $\tilde{R} / v$, for virial velocity $v^2 = GM/\tilde{R}$, also scales as $1/H (a_c)$; see also \cite{Eggemeier:2022gyo}.
} 
A semi-analytic calculation of the GW spectrum produced by the initial stage of collapse was carried out in Ref.~\cite{Dalianis:2020gup} using the Zel'dovich approximation.  The frequency spectrum we find differs from that found in Ref.~\cite{Dalianis:2020gup}, but this should not be surprising, as our numerical calculation includes contributions from shell-crossing and beyond.
Our calculations also differ from the results in~Ref.~\cite{Eggemeier:2022gyo}, which, following Ref.~\cite{Jedamzik:2010hq}, separates the gravitational wave emission from matter into contributions from initial collapse and post-collapse mergers and vibrations.  Here the estimated post-collapse contribution to $\Omega_{\mathrm{GW}}$ carries an additional parametric dependence on $(a_\mathrm{c}/a_\mathrm{eq})^p$, where the exponent $p$ is fitted to Ref.~\cite{Eggemeier:2022gyo}'s numerical data.  This post-collapse contribution is responsible for introducing a substantial $T_\mathrm{eq}$-dependence in their predictions for the gravitational wave spectrum; our parametrics are inconsistent with this result.

\end{document}